\documentclass{aa}  

\usepackage{graphicx}
\usepackage{color}
\usepackage{txfonts}
\usepackage[colorlinks=true,citecolor=blue,urlcolor=blue,linkcolor=blue]{hyperref}
\usepackage{multicol,multirow}
\begin{document} 

\makeatletter
\def\instrefs#1{{\def\scsep{\def\scsep{,}}\@for\w:=#1\do{\scsep\ref{inst:\w}}}}
\renewcommand{\inst}[1]{\unskip$^{\instrefs{#1}}$}

\title{Spitzer thermal phase curve of WASP-121 b}

\author{G. Morello\inst{chalmers,iac}
\and
Q. Changeat\inst{esa,stsci}
\and
A. Dyrek\inst{cea}
\and
P.-O. Lagage\inst{cea}
\and
J. C. Tan\inst{chalmers,uva}
}

\institute{
\label{inst:chalmers}Department of Space, Earth and Environment, Chalmers University of Technology, SE-412 96 Gothenburg, Sweden
\and
\label{inst:iac}Instituto de Astrof\'isica de Canarias (IAC), 38205 La Laguna, Tenerife, Spain
\and
\label{inst:esa}European Space Agency (ESA), ESA Office, Space Telescope Science Institute (STScI), Baltimore MD 21218, USA.
\and
\label{inst:stsci}Department of Physics and Astronomy, University College London, Gower Street,WC1E 6BT London, United Kingdom
\and
\label{inst:cea}AIM, CEA, CNRS, Universit\'e Paris-Saclay, Universit\'e Paris Diderot, Sorbonne Paris Cit\'e, F-91191 Gif-sur-Yvette, France
\and
\label{inst:uva}Dept. of Astronomy, University of Virginia, Charlottesville, VA 22904, USA
}


 
  \abstract
   {}
   {We analyse unpublished Spitzer observations of the thermal phase-curve of WASP-121 b, a benchmark ultra-hot Jupiter.}
   {We adopted the wavelet pixel-independent component analysis technique to remove challenging instrumental systematic effects in these datasets and we fit them simultaneously with parametric light-curve models. We also performed phase-curve retrievals to better understand the horizontal and vertical thermal structure of the planetary atmosphere.}
   {We measured planetary brightness temperatures of $\sim$2700\,K (dayside) and $\sim$700--1100\,K (nightside), along with modest peak offsets of 5.9$^{\circ} \pm$1.6 (3.6\,$\mu$m) and 5.0$^{\circ}$$_{-3.1}^{+3.4}$ (4.5\,$\mu$m) after mid-eclipse. These results suggest inefficient heat redistribution in the atmosphere of WASP-121 b. The inferred atmospheric Bond albedo and circulation efficiency align well with observed trends for hot giant exoplanets. Interestingly, the measured peak offsets correspond to a westward hot spot, which has rarely been  observed. 
   We also report consistent transit depths at 3.6 and 4.5\,$\mu$m, along with updated geometric and orbital parameters. Finally, we compared our Spitzer results with previous measurements, including recent JWST observations.}
   {We extracted new information on the thermal properties and dynamics of an exoplanet atmosphere from an especially problematic dataset. This study probes the reliability of exoplanet phase-curve parameters obtained from Spitzer observations when state-of-the-art pipelines are adopted to remove the instrumental systematic effects. It demonstrates that Spitzer phase-curve observations provide a useful baseline for comparison with JWST observations, and shows the increase in parameters precision achieved with the newer telescope.}


\maketitle
%

\section{Introduction}

WASP-121 b is an ultra-hot Jupiter (UHJ) orbiting around an F6 V star in $\sim$1.27\,d. Table \ref{tab:sys_params} reports the stellar and planetary parameters taken from its discovery paper \citep{delrez2016}. WASP-121 b has been targeted by many follow-up studies, based on its nature as an exoplanet amenable to characterisations with various observing techniques. It is especially well suited for atmospheric characterisation by both transmission and emission spectroscopy, owing to its high equilibrium temperature and large size. Some researchers have proposed WASP-121 b as a suitable target to further investigate its interior structure and/or shape deformations \citep{akinsanmi2019,hellard2020}.

Shortly after the WASP-121 b discovery, \cite{evans2016} detected the 1.4\,$\mu$m water absorption band from a transit observed with the Hubble Space Telescope (HST) using the Wide Field Camera 3 (WFC3) with the G141 grism, covering 1.1-1.7\,$\mu$m. \cite{evans2017} also detected H$_2$O in emission, along with evidence of a stratosphere, using the same instrument setup to observe the planetary eclipse. Earlier atmospheric models of UHJs predicted temperature inversions to occur due to absorption by metal oxides, such as TiO and VO, in their upper atmospheric layers \citep{hubeny2003,fortney2008}. Small features occurring at the blue edge of the HST/WFC3 spectra of WASP-121 b have been tentatively attributed to TiO and VO \citep{evans2017,tsiaras2018}. Based on subsequent transit observations obtained with the HST/Space Telescope Imaging Spectrograph (STIS), covering 0.3-1.0\,$\mu$m, \cite{evans2018} confirmed the presence of VO, but not TiO, at the terminator of WASP-121 b atmosphere.
\cite{mikal-evans2019} found evidence of H$^{-}$ in the emission spectrum of WASP-121 b taken with HST/WFC3 using the G102 grism (0.8-1.1\,$\mu$m), but retracted the previous claim of VO in emission. \cite{mikal-evans2020} refined H$_2$O detection and VO non-detection in the planet dayside by stacking multiple eclipse observations taken with HST/WFC3 G141.
\cite{salz2019} detected an excess of near-UV absorption (0.20-0.27\,$\mu$m) during three transits of WASP-121 b observed with the Ultraviolet/Optical Telescope (UVOT) onboard the Neil Gehrels Swift Observatory. \cite{sing2019} resolved exospheric \ion{Mg}{ii} and \ion{Fe}{ii} lines in the near-UV transmission spectrum observed with HST/STIS.

\begin{table}
\caption{WASP-121 system parameters.}              
\label{tab:sys_params}      
\centering                                      
\begin{tabular}{lc}          
\hline\hline                        
\multicolumn{2}{c}{Stellar parameters} \\    
\hline                                   
$T_{*,\mathrm{eff}}$ [K] & 6460$\pm$140 \\
$\log{g_*}$ [cgs] & 4.2$\pm$0.2 \\
$[\mathrm{Fe/H}]_*$ [dex] & 0.13$\pm$0.09 \\
$M_*$ [$M_{\odot}$] & 1.35$\pm$0.08 \\
$R_*$ [$R_{\odot}$] & 1.46$\pm$0.03 \\
\hline  
\multicolumn{2}{c}{Planetary parameters} \\
\hline
$M_{\mathrm{p}}$ [$M_{\mathrm{Jup}}$] & 1.18$\pm$0.06 \\
$R_{\mathrm{p}}$ [$R_{\mathrm{Jup}}$] & 1.81$\pm$0.04 \\
$T_{\mathrm{p,eq}}$ [K] & 2358$\pm$52 \\
$a$ [au] & 0.02544$\pm$0.00050 \\
$P$ [day] & 1.2749255$_{-2.5 \times 10^{-7}}^{+2.0 \times 10^{-7}}$ \\
$T_0$ [$\mathrm{HJD}_{\mathrm{TDB}}$] & 2456636.34578$_{-0.00010}^{+0.00011}$ \\
\hline
\end{tabular}
\tablefoot{From \cite{delrez2016}.}
\end{table}

\cite{bourrier2020} and \cite{daylan2021} reported two independent analyses of long-term visible photometry of WASP-121 from the Transiting Exoplanet Survey Satellite (TESS, \citealp{ricker2014}) showing strong phase-curve modulations. Both studies measured a strong day-night temperature contrast and a small offset between the maximum emission and substellar points, suggesting low reflectivity and inefficient heat redistribution in the planetary atmosphere. They also found evidence for a temperature inversion, partly caused by H$^{-}$. \cite{mikal-evans2022} analysed two spectroscopic phase curves observed with HST/WFC3 G141, revealing variations in the H$_2$O feature that correspond to a thermal profile warming (cooling) with altitude in the dayside (nightside). These data are consistent with models predicting thermal dissociation of H$_2$O on the dayside and recombination on the nightside \citep{parmentier2018}.
The James Webb Space Telescope (JWST) has recently observed a full phase curve of WASP-121 b using the Near-InfraRed Spectrograph (NIRSpec) with the G395H grism (2.70-5.15\,$\mu$m). \cite{mikal-evans2023} published their first-look analysis of the JWST/NIRSpec data, the results of which align well with those in the previous literature. 

WASP-121 b has also been the subject of numerous ground-based observing campaigns. \cite{kovacs2019} detected a deep planetary eclipse in the 2MASS K band with A Novel Dual Imaging CAMera (ANDICAM) attached to the 1.3-m telescope of the SMARTS Consortium. Multiple studies based on the high-resolution Doppler spectroscopy technique placed severe upper limits on the possible presence of TiO and VO in gaseous form (e.g. \citealp{merritt2020,hoeijmakers2020}). However, the high-resolution spectra revealed a rich inventory of metals and ions in the WASP-121 b atmosphere, including H$\alpha$, \ion{Na}{i}, \ion{Fe}{i}, \ion{Fe}{ii}, \ion{Cr}{i}, \ion{V}{i}, \ion{Mg}{i}, \ion{Ni}{i}, \ion{Ca}{i}, \ion{Ca}{ii}, \ion{K}{i}, \ion{Li}{i}, \ion{Sc}{ii}, \ion{Ba}{ii}, \ion{Co}{i,} and \ion{Sr}{ii} \citep{ben-yami2020,cabot2020,hoeijmakers2020,borsa2021,merritt2021,azevedo_silva2022}.

In this paper, we present the first analysis of two phase curves of WASP-121 b observed with the Spitzer/InfraRed Array Camera (IRAC) channels 1 and 2 \citep{fazio2004}, which operate in photometric passbands centred at 3.6 and 4.5\,$\mu$m, respectively. These data were acquired by the end of January 2018 for the Spitzer program ID 13242 (PI: Tom Evans). Using the wavelet pixel-independent component analysis (ICA) pipeline \citep{morello2016}, we overcome the issues of strong instrumental systematic effects that may have prevented their publication so far. We validate the robustness of our results, comparing them with those from recent JWST observations in similar passbands.

Section \ref{sec:observations} presents the Spitzer/IRAC observations of WASP-121 b phase curves. Section \ref{sec:method} describes the procedure adopted in this work to analyse the data. Section \ref{sec:results} reports our results, including the transit and phase-curve parameters, and derived atmospheric properties. Section \ref{sec:retrievals} discusses the atmospheric properties of WASP-121 b with more details, including the results of phase-curve retrievals. Section \ref{sec:comparisons} compares our results with those from previous observations to obtain a more complete picture of the atmosphere of WASP-121 b, and puts them in the context with other UHJs. Section \ref{sec:conclusions} summarizes the conclusions of our study.

\section{Observations}
\label{sec:observations}

We analysed two phase curves of WASP-121 b observed with Spitzer/IRAC for the program ID 13242 (PI: Tom Evans). Each visit consists of four consecutive astronomical observation requests (AORs) spanning approximately 39 hr, including one transit and two eclipse events. The observations were taken using IRAC sub-array readout mode with 2\,s frame time \citep{fazio2004}. In this mode, 64 frames are taken consecutively, with a delay of 1.27\,s after reset. In total, 69,184 frames were acquired per visit, split unequally across the four AORs, but analogously for the two visits. The first visit made use of IRAC channel 2, that is, photometric filter with $\sim$4.0-5.0\,$\mu$m passband and effective wavelength of 4.5\,$\mu$m. The second visit made use of IRAC channel 1, that is, photometric filter with $\sim$3.2-3.9\,$\mu$m passband and effective wavelength of 3.6\,$\mu$m. Table \ref{tab:obs} summarises the main details of the observations.

\section{Data analysis}
\label{sec:method}

\begin{table*}
\caption{Spitzer/IRAC datasets analysed for this study.}    
\label{tab:obs}      
\centering                                      
\begin{tabular}{ccccccc}          
\hline\hline
Filter\tablefootmark{i} & Prog. ID & AORs\tablefootmark{ii}  & UT start date\tablefootmark{iii} & N$_\mathrm{frames}$ & Mode\tablefootmark{iv} & Pip.\tablefootmark{v} \\
\hline
Ch1 (3.6 $\mu$m) & 13242 & 64973056 & 2018-01-29 16:47:26 & 17536 & sub, 2.0 & S19.2.0 \\
 & " & 64974080 & 2018-01-30 02:42:12 & 9408 & " & " \\
 & " & 64973568 & 2018-01-30 08:03:03 & 21120 & " & " \\
 & " & 64975104 & 2018-01-30 19:58:36 & 21120 & " & " \\
\hline
Ch2 (4.5 $\mu$m) & 13242 & 64974592 & 2018-01-27 03:53:22 & 17536 & sub, 2.0 & S19.2.0 \\
 & " & 64972544 & 2018-01-27 13:48:00 & 9408 & " & " \\
 & " & 64974848 & 2018-01-27 19:08:43 & 21120 & " & " \\
 & " & 64972288 & 2018-01-28 07:04:09 & 21120 & " & " \\
\hline
\end{tabular}
\tablefoot{
\tablefoottext{i}{IRAC channel and central wavelength.}
\tablefoottext{ii}{Astronomical observation requests.}
\tablefoottext{iii}{Timestamp of the first frame of the AOR.}
\tablefoottext{iv}{Readout mode and frame time in seconds.}
\tablefoottext{v}{Pipeline version of the basic calibrated data.}}
\end{table*}

\begin{figure*}
\centering
\includegraphics[width=0.95\hsize]{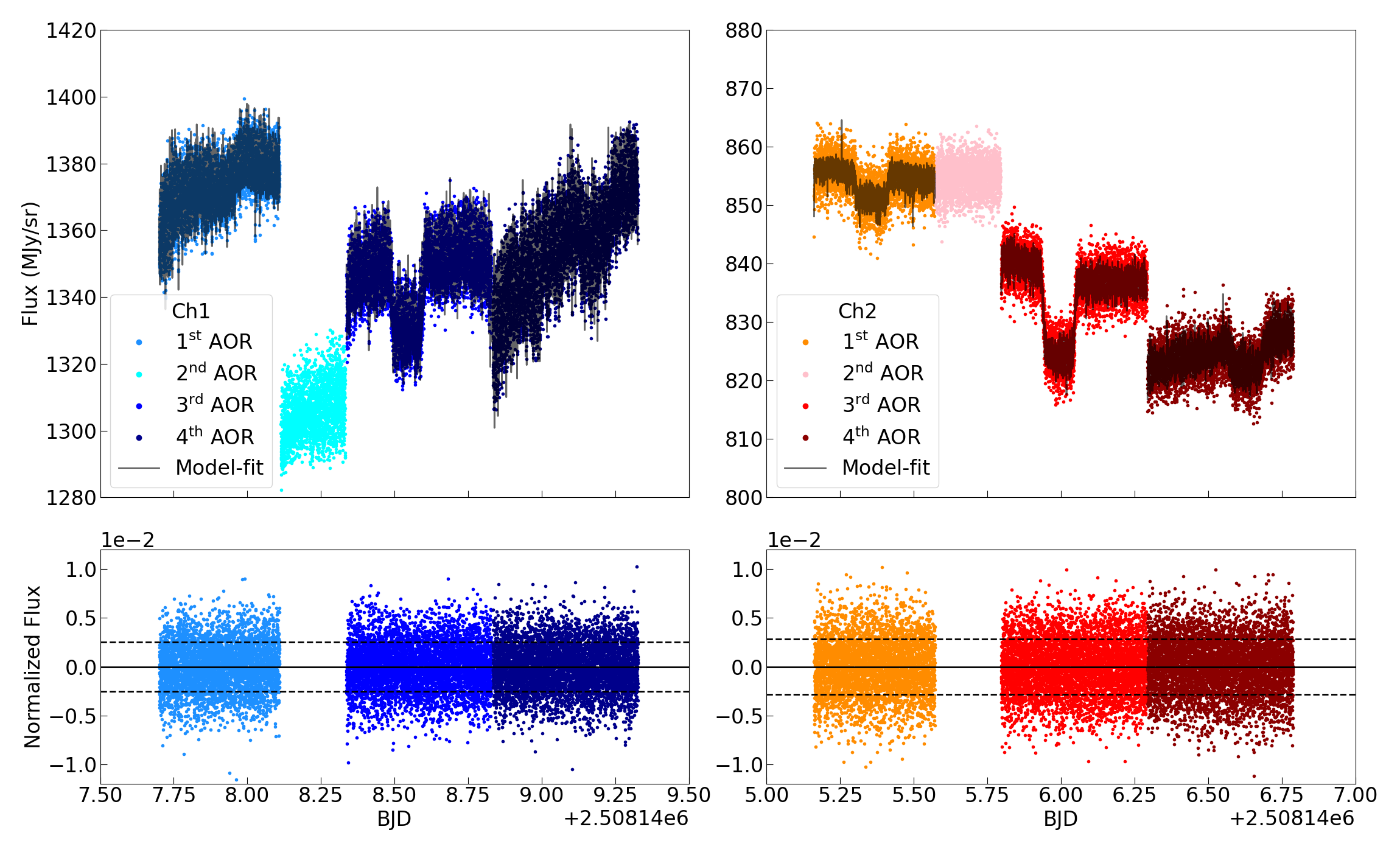}
\caption{Raw light curves obtained for the Spitzer/IRAC observations at 3.6\,$\mu$m (left panel) and 4.5\,$\mu$m (right panel), shown at the top and represented by dots of different colours for each AOR. The relevant best-fit models, including instrumental systematics, are overplotted with solid lines. Note that the second AOR of each visit was discarded from the light-curve fits. The bottom panels show
residuals from the above light curves and models (dots), and standard deviations (black lines).}
\label{fig:lc_raw}
\end{figure*}

\subsection{Raw photometry extraction}
\label{sec:raw_phot}

We downloaded the basic calibrated data (BCD, files extension `\_bcd.fits') from the Spitzer Heritage Archive \citep{wu2010}. The BCD are flat-fielded and flux-calibrated frames \citep{fazio2004,IRAChandbook}. We extracted the pixel time series from 5$\times$5 arrays where the central pixel records the highest flux in most frames within an AOR. The raw light curves were computed as the sum of pixel time series from the 5x5 arrays. We note that the selected array could vary between AORs within the same visit. We also attempted to use a single array for each entire visit, but this choice increased the photometric offsets between AORs and degraded the performance of our data detrending method.

We flagged and corrected outliers in the raw light curves through the following procedure. First, we computed the smoothed reference light curves as the sliding window medians of binned raw light curves. We adopted a bin size of 64, corresponding to the original data cube size, and a sliding window size of five. Second, we computed the reference noise level for the unbinned raw light curves as the median of the moving standard deviation with a sliding window of five. Third, we identified outliers as those points that are more than 5$\sigma$ away from the smoothed references. Fourth, we replaced the sets of consecutive outliers with the vector means of the adjacent sets, or, equivalently, via a linear interpolation in the case of isolated outliers. The replacements were applied to all pixel light curves, and not just to the raw light curves. We iterated the third and fourth steps until there were no outliers left in the raw light curves.

Finally, we binned all the light curves by a factor of four, corresponding to an integration time of 8\,$s$, to speed up the following data analysis. The chosen bin size is a conservative one, being much smaller than the occultation timescales \citep{kipping2010binning}. We could have adopted a larger bin size for the parts outside the occultations, based on the phase curve timescales, but our choice also minimizes the impact of correlated noise \citep{morello2022pcbin}. Figure \ref{fig:lc_raw} shows the binned raw light curves analysed in this work.

\begin{figure*}
\centering
\includegraphics[width=0.95\hsize]{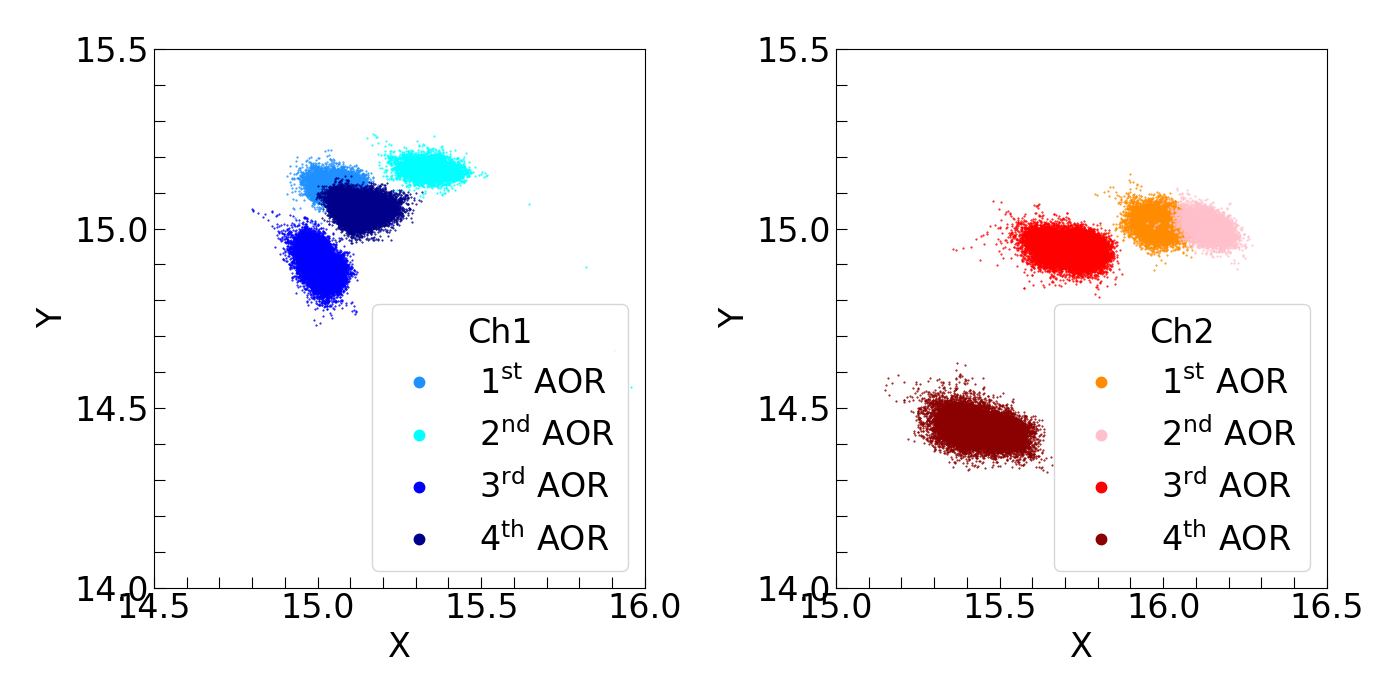}
\caption{Centroid coordinates of the Spitzer/IRAC images, using different colours for each AOR (as in Figure \ref{fig:lc_raw}). Note: there are pointing offsets of few tenths of pixels between consecutive AORs.}
\label{fig:centroids}
\end{figure*}

For illustrative purposes only, we calculated the coordinates of the stellar centroid using the centre-of-light method, as implemented by \cite{morello2015single}. Figure \ref{fig:centroids} shows the $x$ and $y$ coordinates obtained for both visits. It appears by eye that the centroids describe different loci in the $x$-$y$ plane for each AOR. Pointing is stable mostly within 1-2 tenths of the pixel side during an AOR, then jumps abruptly by up to more than half a pixel when starting a new AOR. The larger-than-usual discontinuities likely make these datasets especially challenging to analyse compared to other Spitzer/IRAC phase-curve observations (e.g. \citealp{stevenson2017,morello2019}).

\subsection{Data detrending}

We applied the wavelet pixel-ICA technique, which is one of the most efficient for detrending Spitzer/IRAC time series \citep{ingalls2016}. ICA is a blind source separation technique with a wide range of applications, including many astrophysical fields (e.g. \citealp{maino2002,maino2007,wang2010,wang2013,chapman2012,waldmann2012,waldmann2014,waldmann2013,damiano2017,rodriguez-montoya2018,di_marcantonio2019}). It performs a linear transformation of input mixed signals into maximally independent components \citep{hyvarinen2001}. The wavelet pixel-ICA technique uses wavelet-transformed pixel light curves as input for the ICA \citep{morello2016}. It is an improvement on the pixel-ICA technique, which instead used pixel light curves in the time domain \citep{morello2014,morello2015,morello2015single}. As in previous papers, here we applied  a single-level discrete wavelet transform (DWT) to the pixel light curves, but adopting the Haar wavelet \citep{haar1910} instead of the more complex Daubechies-4 one \citep{daubechies1992}. We checked, however, that the choice of wavelet function does not noticeably affect the ICA transform. The adopted ICA algorithm is MULTICOMBI \citep{tichavsky2008}, as always. This time we wrapped the original MATLAB source code for use in a Python script.

We initially tried to concatenate the pixel light curves from multiple AORs to form a single set of input signals to be transformed with ICA for each visit. \cite{morello2019} successfully adopted this approach to detrend Spitzer/IRAC phase curves of WASP-43 b. The same approach failed on the WASP-121 b data presented here, most likely due to larger pointing jumps between consecutive AORs. Therefore, we decided to perform individual ICA transforms for each AOR. We excluded the second AORs from each visit. In fact, these AORs do not contain an astrophysical signal with a well recognizable shape, such as a transit or an eclipse. The lack of morphology makes it difficult to separate the astrophysical component from the instrumental ones.
For all other AORs, we identified the first ICA component as the astrophysical one, containing a clear transit or eclipse signal. Following the usual procedure, these astrophysical components were discarded from the light-curve fits, being replaced by an astrophysical light-curve model. The other 24 components of each AOR were attributed to instrumental systematic signals.

\subsection{Light-curve models}
\label{sec:pc_model}
Our phase-curve model approximates the planetary flux with a double sinusoid,
\begin{equation}
\label{eqn:pc_def}
F_{\mathrm{p}} = c_0 + c_1 \cos{ \left [2 \pi \left ( \phi' - \phi'_1 \right ) \right ]} + c_2 \cos{ \left [4 \pi \left ( \phi' - \phi'_2 \right ) \right ]},
\end{equation}
where $\phi' = \phi - \Delta \phi_{\mathrm{ltd}}$ is the orbital phase corrected for the light travel delay. The orbital phase is
\begin{equation}
\phi = \frac{t - T_0}{P} - n,
\end{equation}
where $T_0$ is the epoch of transit, $P$ is the orbital period, and $n$ is an integer number usually chosen such that $-1 \le \phi \le 1$. The light travel delay accounts for the displacements of the planet along the line of sight with respect to inferior conjunction. For a circular orbit,
\begin{equation}
\Delta \phi_{\mathrm{ltd}} = \frac{a \sin{i} \left [ 1 - \cos{( 2 \pi \phi )} \right ]}{c P},
\end{equation}
where $a$ and $i$ are the orbital semimajor axis and inclination, and $c$ is the speed of light.

We adopted \texttt{PYLIGHTCURVE}\footnote{\url{https://github.com/ucl-exoplanets/pylightcurve}} \citep{tsiaras2016} to model the occultations, which is based on the formalism from \cite{pal2008}. We computed the stellar limb-darkening coefficients through \texttt{ExoTETHyS}\footnote{\url{https://github.com/ucl-exoplanets/ExoTETHyS}} \citep{morello2020joss,morello2020}, using spectral model grids from the \texttt{PHOENIX} library \citep{claret2012,claret2013,husser2013} and the so-called claret-4 parametrisation \citep{claret2000}. Table \ref{tab:ldcs} reports the limb-darkening coefficients for both Spitzer/IRAC passbands.

\begin{table}[]
\caption{Limb-darkening coefficients for WASP-121.}
\centering
\begin{tabular}{ccc}
\hline\hline
Exponent\tablefootmark{i} & Ch1 (3.6\,$\mu$m) & Ch2 (4.5\,$\mu$m) \\
\hline
\noalign{\smallskip}
1/2 & 0.354241 & 0.341362 \\
1 & -0.134867 & -0.211578 \\
3/2 & 0.068047 & 0.150017 \\
2 & -0.017143 & -0.045761 \\
\hline
\end{tabular}
\tablefoot{
\tablefoottext{i}{of the power term from the four-coefficient law \citep{claret2000}.}
}
\label{tab:ldcs}
\end{table}

\subsection{Data fitting}

\begin{table}[]
\caption{Prior probability distributions of the fitted parameters.}
\centering
\begin{tabular}{ll}
\hline\hline
Parameter [units] & Prior \\
\hline
\noalign{\smallskip}
$p$ [] \tablefootmark{i} & $\mathcal{U}(0,1)$ \\
$P$ [d] \tablefootmark{ii} & $\mathcal{N}(1.2749255,2.5 \times 10^{-7})$ \\
$T_0$ [$\mathrm{HJD}_{\mathrm{TDB}}$] \tablefootmark{iii} & $\mathcal{U}(2456635.54895,2456635.86769)$ \\
$b$ [] \tablefootmark{iv} & $\mathcal{U}(0,1)$ \\
$T_{14}$ [hr] \tablefootmark{v} & $\mathcal{U}(0,8.6616)$ \\
$a$ [au] \tablefootmark{vi} & $\mathcal{N}(0.02544,0.00050)$ \\
$c_0$ [] \tablefootmark{vii} & $\mathcal{U}(0,1)$ \\
$c_1$ [] \tablefootmark{vii} & $\mathcal{U}(-1,1)$ \\
$\phi'_1$ [] \tablefootmark{vii} & $\mathcal{U}(-0.5,0.5)$ \\
$c_2$ [] \tablefootmark{vii} & $\mathcal{U}(-1,1)$ \\
$\phi'_2$ [] \tablefootmark{vii} & $\mathcal{U}(-0.5,0.5)$ \\
$N_{\mathrm{AOR}}$ [MJy/sr] \tablefootmark{viii} & $\mathcal{U}^{3}(0,3000)$ \\
$k_{\mathrm{ICA}}$ [] \tablefootmark{ix} & $\mathcal{U}^{72}(-100,100)$ \\
\hline
\end{tabular}
\tablefoot{
$\mathcal{U}(a,b)$ denotes a uniform prior delimited by $a$ and $b$; $\mathcal{N}(\mu,\sigma)$ denotes a normal prior with $\mu$ mean and $\sigma$ width; literature values were taken from \cite{delrez2016}, alias D16. Parameters:
\tablefoottext{i}{planet/star radius ratio, $p = R_{\mathrm{p}}/R_*$;}
\tablefoottext{ii}{orbital period, normal prior from D16;}
\tablefoottext{iii}{epoch of transit, uniform prior interval centred on D16 value with width equal to 0.25 $P$;}
\tablefoottext{iv}{impact parameter, $b = a \cos{i} / R_*$;}
\tablefoottext{v}{total transit duration from first to fourth contact, uniform prior interval from 0 to 3$\times$ D16 value;}
\tablefoottext{vi}{orbital semimajor axis, normal prior from D16;}
\tablefoottext{vii}{phase-curve parameters defined in Equation \ref{eqn:pc_def};}
\tablefoottext{viii}{scaling factors for the light-curve model, independent for each AOR;}
\tablefoottext{ix}{scaling factors for the ICA components.}
}
\label{tab:priors}
\end{table}

We performed similar independent fits on both visits, taken separately.
For each visit, we simultaneously fitted the light-curve model and instrumental systematic effects to the raw light curve, discarding the second AOR. For each AOR segment considered, we fit a linear combination of the light-curve model and 24 ICA components attributed to instrumental signals. The scaling factors for the light-curve model and the ICA components of different AORs were independent parameters. The astrophysical parameters were planet-to-star radius ratio ($p$), orbital period ($P$), epoch of transit ($T_0$), impact parameter ($b$), total transit duration ($T_{14}$), orbital semi-major axis ($a$), and five phase-curve parameters (as in Equation \ref{eqn:pc_def}). These parameters were shared among the AORs of the same visit. 

Table \ref{tab:priors} reports the Bayesian priors assigned to the 86 free parameters listed above. We set large uniform priors for almost all parameters. The orbital period and semi-major axis are wavelength-independent parameters that are very well known from previous observations, for which we adopted normal priors based on the results from \cite{delrez2016}. 
We performed a preliminary optimisation using \texttt{scipy.optimize.minimize} with the Nelder-Mead method \citep{nelder1965}. The root mean square (rms) of the corresponding residuals was assigned as the error bar to each photometric point, which is typically larger than the nominal error bars. Then we ran emcee \citep{emceev3joss} with 300 walkers and 200,000 iterations. Each walker
was initialised with a random value close to the preliminary parameter estimate.
The first 50,000 iterations were discarded as burn-in.

\subsection{Alternative fits}
\label{sec:alt_fits}
We tested fitting a phase-curve model with a single sinusoid, namely, fixing $c_2=0$ in Equation \ref{eqn:pc_def}. The corresponding results were not preferred, as explained in Section \ref{sec:aic_bic}. We also tried fixing the geometric and orbital parameters to better constrain the difference in transit depth between IRAC passbands, as discussed in Section \ref{sec:res_params}. These tests were not used to calculate the final parameters reported in Table \ref{tab:posteriors}.

\section{Results}
\label{sec:results}

\begin{figure}
\includegraphics[width=0.95\hsize]{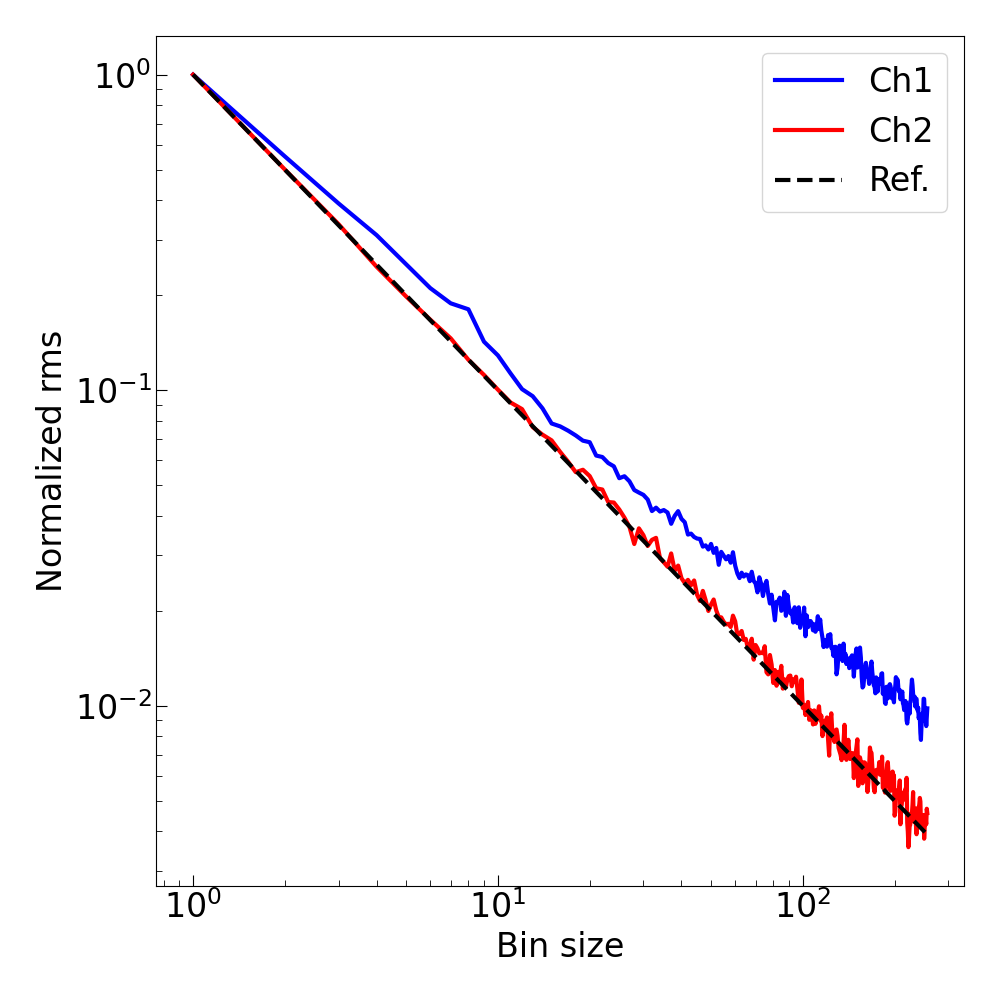}
\caption{Normalized rms of residuals as function of bin size for the 3.6\,$\mu$m (blue) and 4.5\,$\mu$m (red) visits. The unit bin size is 8\,s. Depending on the bin size, a few points are discarded at the beginning of each AOR to avoid bins over the gaps between AORs. The black dashed line shows the theoretical behavior for Gaussian residuals.}
\label{fig:rms_binning}
\end{figure}

Figure \ref{fig:lc_raw} shows the best-fit models to the raw light curves and the corresponding residuals. The rms amplitudes of the normalised residuals are 2.45$\times$10$^{-3}$ for the 3.6\,$\mu$m visit, and 2.81$\times$10$^{-3}$ for the 4.5\,$\mu$m visit. We estimated them to be $\sim$28.7\% and 6.8\% above the photon noise limit. Figure \ref{fig:rms_binning} shows the rms amplitudes of the binned residuals versus the bin size. The 4.5\,$\mu$m residuals show no significant deviations from the theoretical behaviour of white noise. The 3.6\,$\mu$m residuals present a modest amount of correlated noise, as is often the case for observations with this IRAC channel (e.g. \citealp{maxted2013,stevenson2017,zhang2018,dang2022}).

\subsection{Model selection}
\label{sec:aic_bic}
We compared the phase-curve models with a single or double sinusoid, as described in Sections \ref{sec:pc_model}--\ref{sec:alt_fits}. We considered the Bayesian information criterion (BIC, \citealp{schwarz1978}) and the Akaike information criterion (AIC, \citealp{akaike1974}) to guide our model selection.
For the 3.6\,$\mu$m observation, the double sinusoid is statistically preferred according to both criteria with $\Delta \mathrm{BIC}=-52$ and $\Delta \mathrm{AIC}=-67$. We note that $| \Delta \mathrm{BIC} | > 10$ (or $| \Delta \mathrm{AIC} | > 10$) indicates a very strong evidence in favour of either model, based on the scale by \cite{raftery1995}. For the 4.5\,$\mu$m observation, the analogous differences have opposite sign, $\Delta \mathrm{BIC}=18$ and $\Delta \mathrm{AIC}=2.7$, thus favouring the single sinusoid model. We note that $| \Delta \mathrm{AIC} | \sim 2$ indicates a weak statistical preference for the model with the lowest AIC.

Finally, we selected the results obtained with the double sinusoid model for both observations. The choice to adopt the same parametrisation for both observations was taken to ensure homogeneity in their analyses. In fact, we expect the same physical phenomena to be present in observations of the same system at multiple wavelengths, albeit with different relative amplitudes and/or signal-to-noise ratios (S/Ns). Additionally, the double sinusoid parametrisation is more flexible and includes the single sinusoid as a special subcase. Even if the second sinusoid were superfluous to reproduce the 4.5\,$\mu$m data, the fit should find a null amplitude ($c_2 \sim 0$) without biasing the other parameters.

Indeed, the fits with single and double sinusoid led to 1$\sigma$ consistent results for the 4.5\,$\mu$m observation, and slightly more conservative error bars by up to $\sim$10\% when using the more complete parametrisation. As expected, the differences between the two sets of results for the 3.6\,$\mu$m observation are more significant, sometimes exceeding the 3$\sigma$ level. The single sinusoid led to unphysical results for the 3.6\,$\mu$m phase curve, such as a negative nightside flux within $\sim$2$\sigma$. This issue is overcome by adopting the double sinusoid model.

\subsection{Transit and phase-curve parameters}
\label{sec:res_params}

\begin{table*}[]
\caption{Posterior distributions of the fitted and derived parameters.}
\centering
\begin{tabular}{lcc}
\hline\hline
Parameter [Units] & Ch1 (3.6\,$\mu$m) & Ch2 (4.5\,$\mu$m) \\
\hline
\noalign{\smallskip}
\multicolumn{3}{c}{Fitted} \\
$p$ [] & 0.1228$\pm$0.0005 & 0.1231$\pm$0.0005 \\
$P$ [d] & 1.2749255$\pm$2.5$\times$10$^{-7}$ & 1.2749255$\pm$2.5$\times$10$^{-7}$ \\
$T_0$ [$\mathrm{HJD}_{\mathrm{TDB}}$] & 2456635.7066$\pm$0.0003 & 2456635.7065$\pm$0.0003 \\
$b$ [] & 0.24$_{-0.10}^{+0.07}$ & 0.15$_{-0.10}^{+0.09}$ \\
$T_{14}$ [hr] & 2.926$_{-0.014}^{+0.016}$ & 2.909$_{-0.011}^{+0.014}$ \\
$a$ [au] & 0.02544$\pm$0.00050 & 0.02544$\pm$0.00050 \\
\noalign{\smallskip}
\hline
\noalign{\smallskip}
\multicolumn{3}{c}{Derived (transit)} \\
$p^2$ [$\times$10$^{-2}$] & 1.508$\pm$0.012 & 1.514$\pm$0.013 \\
$a_0$ [] & 3.71$\pm$0.07 & 3.79$_{-0.06}^{+0.03}$ \\
$i$ [deg] & 86.2$_{-1.2}^{+1.6}$ & 87.8$\pm$1.4 \\
\noalign{\smallskip}
\hline
\noalign{\smallskip}
\multicolumn{3}{c}{Derived (phase curve)} \\
$F_{\mathrm{day}}^{\mathrm{MAX}}$ [$\times$10$^{-3}$] & 4.23$\pm$0.08 & 5.09$\pm$0.09 \\
$F_{\mathrm{night}}^{\mathrm{MIN}}$ [$\times$10$^{-3}$] & 0.05$\pm$0.24 & 0.71$\pm$0.27 \\
$\Delta \phi_{\mathrm{day}}^{\mathrm{MAX}}$ [deg] & 5.9$\pm$1.6 & 5.0$_{-3.1}^{+3.4}$ \\
$\Delta \phi_{\mathrm{night}}^{\mathrm{MIN}}$ [deg] & -31$\pm$6 & 4.5$\pm$2.9 \\
$T_{\mathrm{day}}^{\mathrm{MAX}}$ [K] & 2670$_{-40}^{+55}$ & 2700$_{-50}^{+70}$ \\
$T_{\mathrm{night}}^{\mathrm{MIN}}$ [K] & 710$_{-710}^{+270}$ & 1130$_{-160}^{+130}$ \\
$A_{\mathrm{b}}$ [] & 0.37$_{-0.09}^{+0.07}$ & 0.32$_{-0.10}^{+0.08}$ \\
$\varepsilon$ [] & 0.013$_{-0.013}^{+0.034}$ & 0.077$_{-0.034}^{+0.040}$ \\
\noalign{\smallskip}
\hline
\noalign{\smallskip}
$F(\phi'=0.5)$ [$\times$10$^{-3}$] & 4.21$\pm$0.08 & 5.08$\pm$0.09 \\
$F(\phi'=0)$ [$\times$10$^{-3}$] & 0.22$\pm$0.26 & 0.72$\pm$0.27 \\
$\bar{T}_{\mathrm{p}}(\phi'=0.5)$ [K] & 2665$_{-40}^{+55}$ & 2700$_{-50}^{+70}$ \\
$\bar{T}_{\mathrm{p}}(\phi'=0)$ [K] & 810$_{-810}^{+360}$ & 1130$_{-160}^{+130}$ \\
\noalign{\smallskip}
\hline
\end{tabular}
\label{tab:posteriors}
\end{table*}

Table \ref{tab:posteriors} reports the best-fit parameters and others derived from those. The corresponding corner plots are presented in Appendix \ref{app:cornerplots}. The transit geometric and orbital parameters estimated independently from the two observations are consistent within 1$\sigma$. There is no evidence of different transit depths ($p^2$) at 3.6 and 4.5\,$\mu$m within their error bars of 120-130 ppm. We also attempted to fit both light curves with fixed geometric and orbital parameters ($b$, $P$ and $a$) to reduce their degeneracies with transit depth. When fixing the above parameters, the differential transit depth slightly increased from 60 to 100 ppm, which is still below the 1$\sigma$ error bars. 

From the posterior distributions of the phase-curve coefficients, we numerically calculated the dayside maximum and nightside minimum fluxes ($F_{\mathrm{day}}^{\mathrm{MAX}}$ and $F_{\mathrm{night}}^{\mathrm{MIN}}$) and their phase offsets from mid-transit and mid-eclipse time ($\Delta \phi_{\mathrm{day}}^{\mathrm{MAX}}$ and $\Delta \phi_{\mathrm{night}}^{\mathrm{MIN}}$). We used the \texttt{ExoTETHyS.BOATS} subpackage \citep{morello2021} to determine the brightness temperatures corresponding to the measured planetary fluxes. We obtained $F_{\mathrm{day}}^{\mathrm{MAX}} = (4.23 \pm 0.08 ) \times 10^{-3}$ and $(5.09 \pm 0.09 ) \times 10^{-3}$ at 3.6 and 4.5\,$\mu$m, corresponding to similar brightness temperatures of $T_{\mathrm{day}}^{\mathrm{MAX}} = 2670_{-40}^{+55}$ and $2700_{-50}^{+70}$\,K, respectively. The two phase-curve maxima occur with slight offsets after mid-eclipse, $\Delta \phi_{\mathrm{day}}^{\mathrm{MAX}} = 5^{\circ}.9 \pm 1^{\circ}.6$ and $5^{\circ}.0_{-3.4}^{+3.1}$ at 3.6 and 4.5\,$\mu$m, respectively. We could only place an upper limit of $F_{\mathrm{night}}^{\mathrm{MIN}} = ( 0.05 \pm 0.24 ) \times 10^{-3}$ on the 3.6\,$\mu$m nightside minimum flux, corresponding to $T_{\mathrm{night}}^{\mathrm{MIN}} = 710_{-710}^{+270}$\,K. We also measured $F_{\mathrm{night}}^{\mathrm{MIN}} = ( 0.71 \pm 0.27 ) \times 10^{-3}$ at 4.5\,$\mu$m, corresponding to $T_{\mathrm{night}}^{\mathrm{MIN}} = 1130_{-160}^{+130}$\,K.
Following the formulation by \cite{cowan2011}, we estimated the Bond albedo ($A_{\mathrm{b}}$) and circulation efficiency ($\varepsilon$) from the brightness temperatures. We obtained $A_{\mathrm{b}} = 0.37_{-0.09}^{+0.07}$ and $\varepsilon = 0.013_{-0.013}^{+0.034}$ (3.6\,$\mu$m), and $A_{\mathrm{b}} = 0.32_{-0.10}^{+0.08}$ and $\varepsilon = 0.077_{-0.034}^{+0.040}$ (4.5\,$\mu$m).

\section{Discussion}
\label{sec:discussion}

\begin{figure*}
\centering
\includegraphics[width=0.49\textwidth]{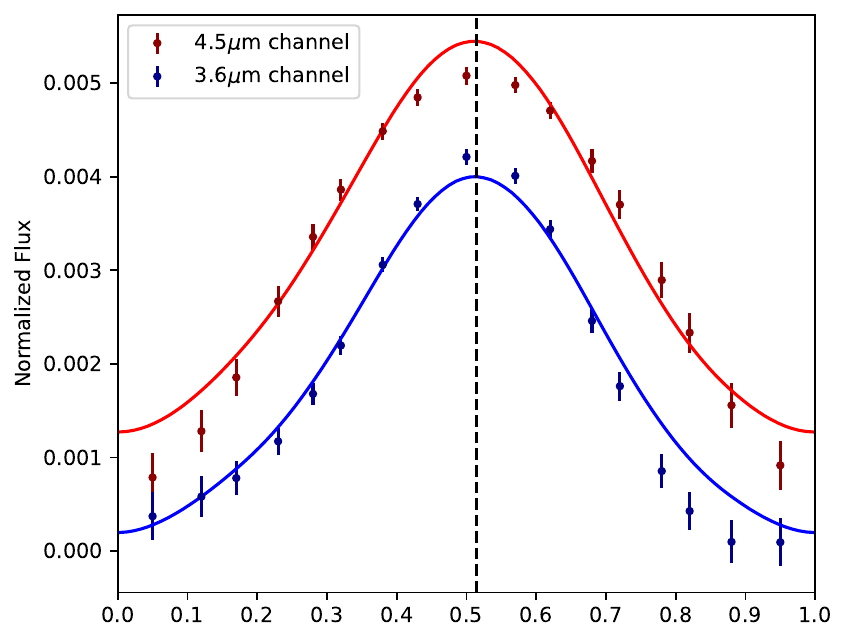}
\includegraphics[width=0.49\textwidth]{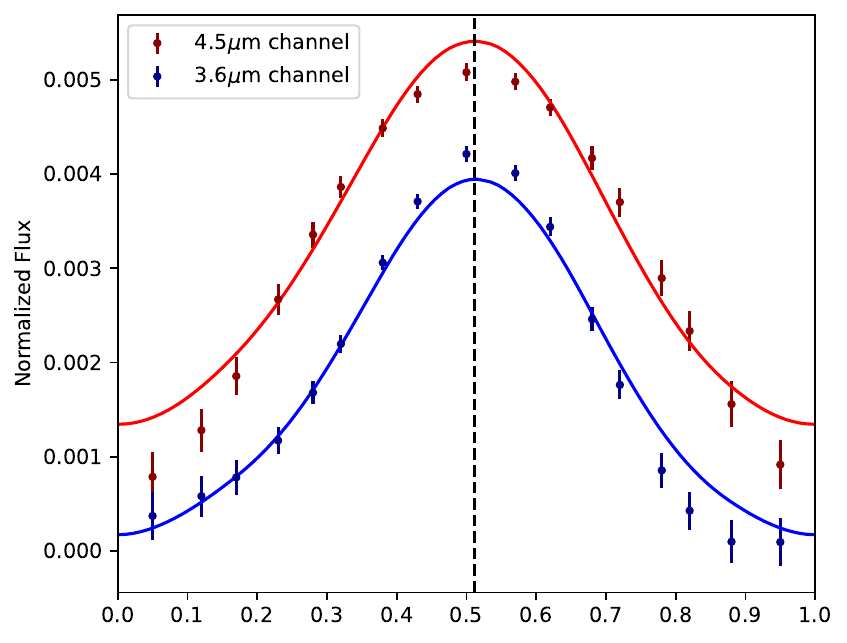}
\includegraphics[width=0.49\textwidth]{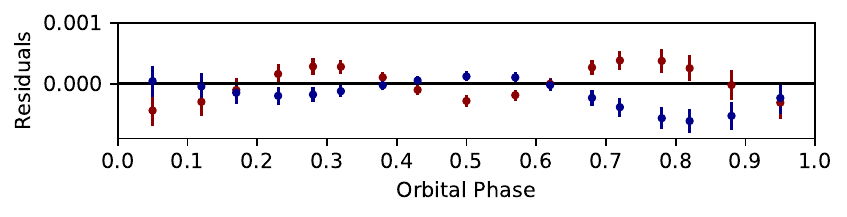}
\includegraphics[width=0.49\textwidth]{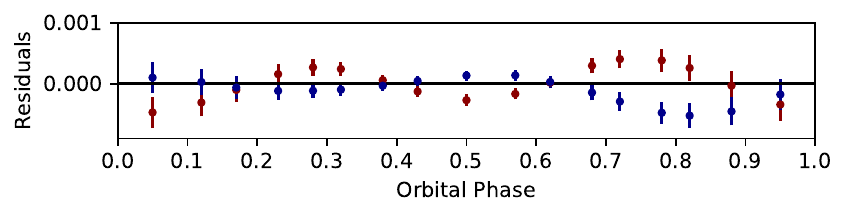}

\caption{Parametric Spitzer phase curves with error bars based on Equation \ref{eqn:pc_def} and best-fit models from our 1.5D phase-curve retrieval. The dashed vertical lines indicate the position of the hot spot's centre. Note: the orbital phase has been corrected for the light travel delay, so that mid-eclipse occurs at $\phi'=$0.5. Left: Chemistry is fixed at $Z_{\mathrm{p}} = 1 \, Z_\odot$; Right: Chemistry is fixed at $Z_{\mathrm{p}} = 10 \, Z_\odot$. }
\label{fig:retrieval_pc}
\end{figure*}

\subsection{WASP-121 b atmosphere overview}
\label{sec:retrievals}
The dayside emission spectrum of WASP-121 b, limited to the Spitzer/IRAC 3.6 and 4.5\,$\mu$m passbands, is consistent with that of a blackbody at 2680$_{-45}^{+60}$\,K (weighted average). The nightside emission is also consistent with that from a blackbody with 1100$_{-220}^{+165}$\,K. From this data, there is no evidence of molecular species either in emission (blackbody spectra) or in transmission (constant transit depths). The strong day-night contrast and small peak offsets point towards inefficient heat redistribution in the WASP-121 b atmosphere. Based on the weighted average blackbody temperatures, we report $\varepsilon = 0.07_{-0.04}^{+0.05}$.
An interesting feature from both phase curves is the indication of a westward hot-spot offset, in the opposite direction to that predicted by most global circulation models (GCM) of hot Jupiter atmospheres (e.g. \citealp{showman2002,showman2009,perna2012,mendonca2016,deitrick2020}). Although they are rare, westward hot-spot offsets have been previously reported for a few hot Jupiters \citep{armstrong2016,dang2018,von_essen2020} and predicted theoretically \citep{rogers2014,rogers2017,hindle2019}.

We performed phase-curve retrievals with the phase-curve plugin \citep{changeat2020, Changeat2021} of \texttt{TauREx} 3.1 \citep{al-refaie2021, al-refaie2022}, the latest version of the \texttt{TauREx} software \citep{waldmann2015b,waldmann2015a}.
For the atmosphere, we assumed three homogeneous regions: hot spot, dayside, and nightside. We computed the emitted flux at given phases using a quadrature integration scheme and fit all the phases for both channels in a single run. Each region is described by a plane-parallel atmosphere composed of 100 layers with pressures ranging from 10 to 10$^{-6}$ bar in log scale. In principle, multiwavelength phase-curve observations may constrain the chemistry of exoplanet atmospheres (e.g. \citealp{stevenson2014,kreidberg2018,arcangeli2019}), and thus potentially inform us about their formation and evolution pathways (e.g. \citealp{madhusudhan2014,turrini2021,cevallos_soto2022}). However, the information content in regard to the chemistry is relatively low and is degenerate in Spitzer data, especially if the thermal structure is also unknown, so we coupled the chemistry between the three regions of the planet and assumed chemical equilibrium. In the retrievals, the values for the metallicity ($Z_{\mathrm{p}}$) and the carbon-to-oxygen ($({\rm C/O})_{\mathrm{p}}$) ratio were left fixed. We tested runs with $Z_{\mathrm{p}} = 1-10 \, Z_\odot$ and with $({\rm C/O})_{\mathrm{p}}=0.1-1.0$. The thermal profiles were described using a two-point profile with two freely moving nodes. In this model, the hot-spot region is defined by its location and size. We initially attempted to recover both parameters from the data but this led to nonphysical solutions. This issue may occur because with Spitzer data only, the hot-spot size is degenerate with the thermal structure. A similar behaviour was found and explored in more detail in \cite{Changeat2021,Changeat2022b}, even when the HST data are combined with Spitzer data. We therefore fixed the hot-spot size to 40$^{\circ}$, but left the hot-spot offset as a free parameter. For the nightside region, we modeled clouds using an opaque grey cloud model and fitted for the cloud pressure top deck.
The parameter space of this phase-curve model was explored using the MultiNest algorithm \citep{Feroz2009,Buchner_2016} with 512 live points and an evidence tolerance of 0.5. The priors were chosen to be uninformative, i.e., uniform priors with large bounds. More specifically, the hot-spot offset was allowed to vary between -50$^{\circ}$ and 50$^{\circ}$, the temperature of the T-p nodes between 300\,K and 6000\,K, and the pressures of the T-p nodes as well as the top of the cloud deck were explored on the full extent of the atmosphere.

\begin{figure}
\includegraphics[width=0.24\textwidth]{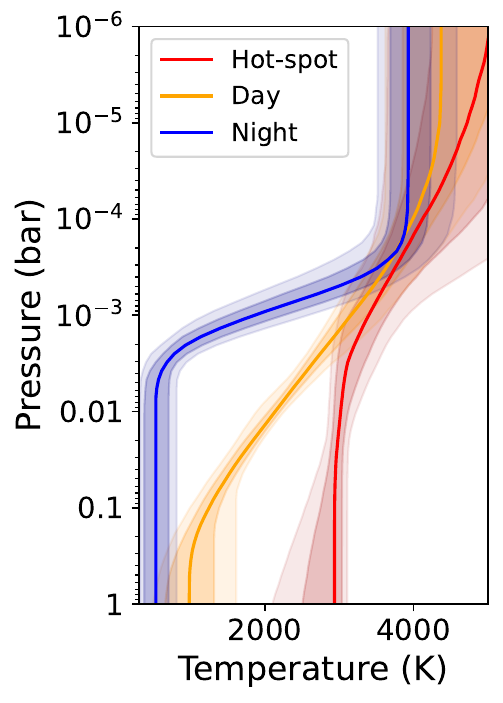}
\includegraphics[width=0.24\textwidth]{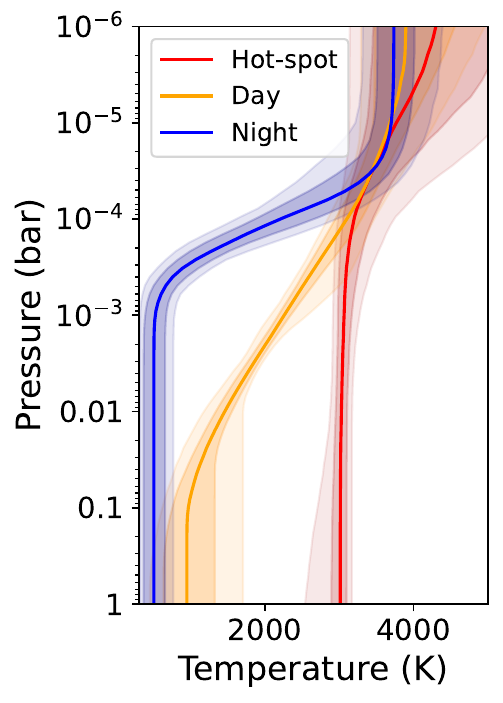}
\caption{Retrieved thermal structures for the $Z_{\mathrm{p}} = 1 \, Z_\odot$ (left) and $Z_{\mathrm{p}} = 10 \, Z_\odot$ (right) phase-curve retrievals, corresponding to the fits in Figure \ref{fig:retrieval_pc}. The shaded regions are the 1 and 3$\sigma$ confidence levels. In both cases, a thermal inversion is needed to explain the data.}
\label{fig:retrieval_tp}
\end{figure}

We show in Figure \ref{fig:retrieval_pc} the Spitzer observations, calculated from the posterior distributions of our parametric fit based on Equation \ref{eqn:pc_def}, and two recovered best-fit atmospheric models. The corresponding retrieved thermal structures are shown in Figure \ref{fig:retrieval_tp}. While both runs indicate the likely presence of a thermal inversion, the altitude of the inversion cannot be inferred from this data as it is degenerate with the chemistry. 

Analysing the posterior distributions (see Figure \ref{fig:corner_atmosphere}) of our atmospheric retrievals, we find that clouds are not required to explain the WASP-121\,b Spitzer data. The hot-spot offset is consistent between the two $Z_{\mathrm{p}} = 1 \, Z_\odot$ and $Z_{\mathrm{p}} = 10 \, Z_\odot$ retrievals, around 9$^{\circ}$ westward.

Looking at the residuals in Figure \ref{fig:retrieval_pc}, we note some discrepancies between our phase-curve models and Spitzer data. The anti-correlated behaviour of 3.6\,$\mu$m and the 4.5\,$\mu$m residuals suggest that the hot-spot offset, size, and temperature (shared in our retrievals) might be different between the two observations. This potential difference in the hot-spot parameters could be a consequence of atmospheric temporal variability or other effects that are not accounted for by our analysis, such as the observations probing different pressure regions, or remaining systematic biases from our data reduction.
We defer further modelling efforts to future work, given the high level of complexity required to reproduce both observations and difficulty in constraining many atmospheric parameters using just two photometric observations.

\begin{figure}
\includegraphics[width=0.95\hsize]{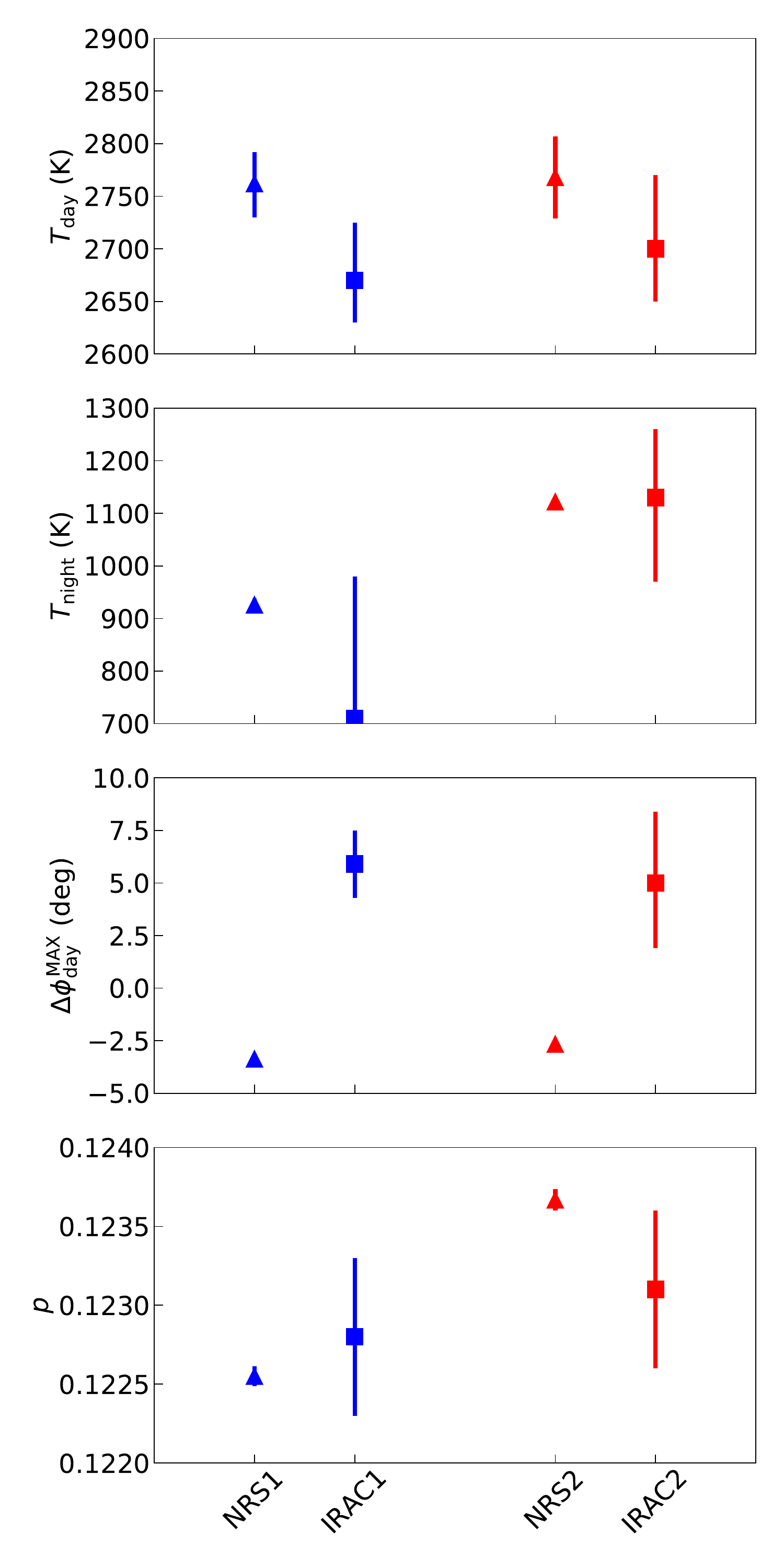}
\caption{Brightness temperatures inferred for the dayside (left panel) and nightside (right panel) of WASP-121 b, shown at the top.  Offsets of the phase-curve maxima relative to conjunction (left panel), and planet-to-star radii ratios (right panel), shown at the bottom. Parameters obtained from the Spitzer/IRAC data are represented with blue and red squares for channel 1 and 2, respectively. Those obtained from the JWST/NIRSpec G395H data are represented with blue and red triangles for NRS1 and NRS2, respectively.}
\label{fig:jwst_comp}
\end{figure}

\begin{figure}
\includegraphics[width=0.95\hsize]{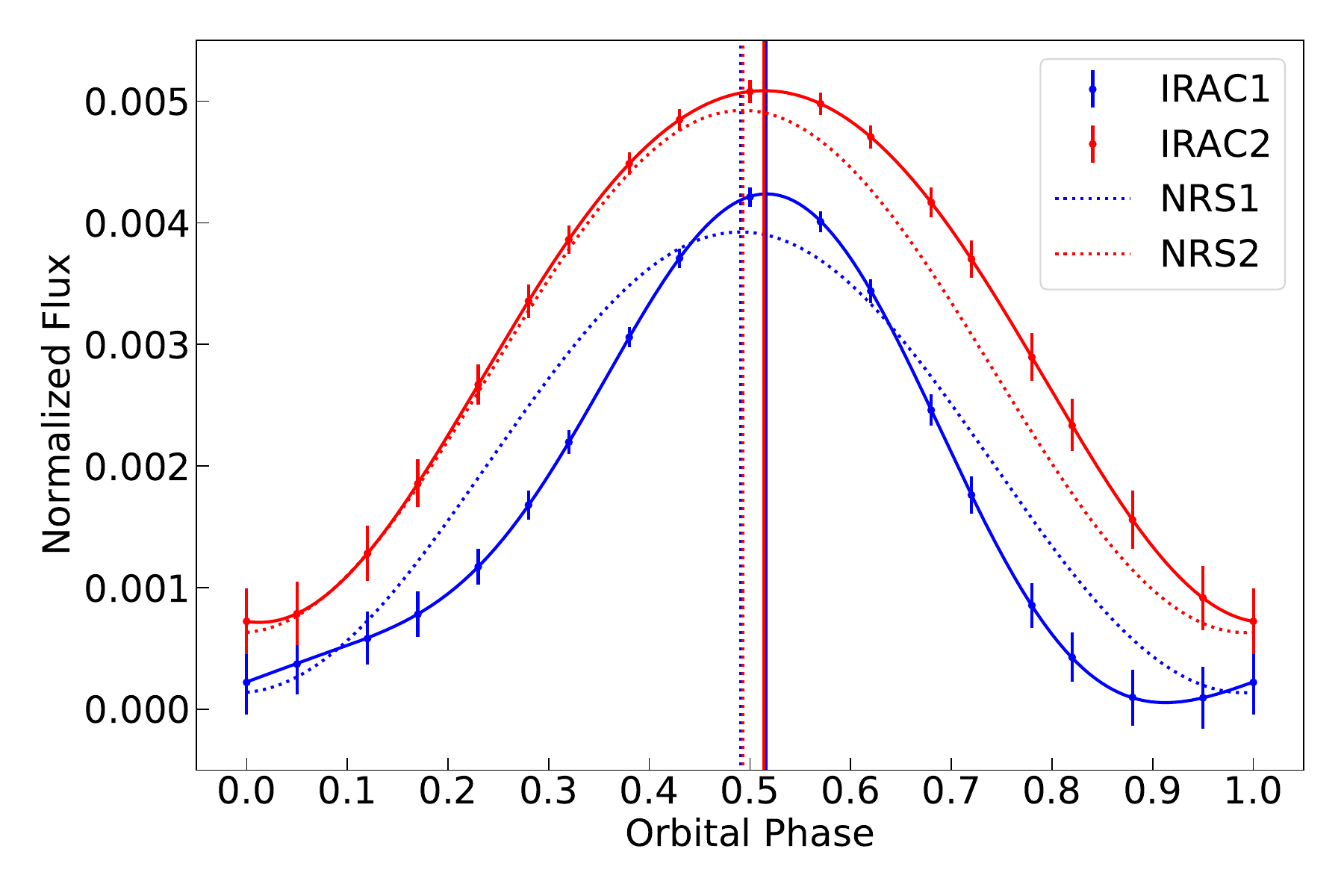}
\caption{Parametric phase-curve profiles for the Spitzer/IRAC observations analysed in this study (solid lines with error bars) and JWST/NIRSpec G395H inferred from \cite{mikal-evans2023} (dotted lines). The vertical lines indicate the position of the maxima.}
\label{fig:jwst_pc_comp}
\end{figure}

\subsection{Comparison with other observations of the same planet}
\label{sec:comparisons}

\subsubsection{JWST/NIRSpec}
The WASP-121 b phase curve was recently observed by JWST/NIRSpec using the G395H grating as part of program GO-1729 (P.I. Mikal-Evans, co-P.I. Kataria). This observing mode makes use of the NRS1 and NRS2 detectors, covering the 2.70-3.72\,$\mu$m and 3.82-5.15\,$\mu$m wavelength ranges. \cite{mikal-evans2023} presented the results of their first look analysis of the broadband light curves, integrated over each detector passband. We note that the NRS1 and NRS2 passbands largely overlap with those of Spitzer/IRAC channels 1 and 2, respectively. Hence, it makes sense to compare the results obtained from Spitzer and JWST observations.

Figure \ref{fig:jwst_comp} shows the comparison between physical parameters from our Spitzer/IRAC data analysis and those based on the JWST/NIRSpec observations reported by \cite{mikal-evans2023}. Figure \ref{fig:jwst_pc_comp} compares the corresponding phase-curve profiles. There is a good agreement between the two sets of parameters, albeit with larger error bars for the Spitzer/IRAC ones.
In particular, the two-points dayside emission spectra of WASP-121 b inferred from Spitzer/IRAC or JWST/NIRSpec are both consistent with that of blackbodies. There is an apparent offset of $\sim$80\,K between the two sets of brightness temperatures, which is not statistically significant.
The nightside temperatures reported for the Spitzer/IRAC and JWST/NIRSpec passbands have similar trends, the bluer temperatures being $\sim$200-400\,K lower than the redder. While the difference for the JWST/NIRSpec passbands is significant at the 12$\sigma$ level, the statistical significance of the difference is decreased by the order-of-magnitude larger error bars for the Spitzer/IRAC measurements.
The phase-curve maxima present different offsets from mid-eclipse, ranging from modest eastward to westward hot-spot positions for JWST/NIRSpec and Spitzer/IRAC measurements, respectively. The reported JWST/NIRSpec offsets are 3.36$^{\circ}\pm$0.11$^{\circ}$ (NRS1) and 2.66$^{\circ}\pm$0.12$^{\circ}$ (NRS2) prior to mid-eclipse.
These differences may reveal the variable weather of WASP-121 b \citep{cho2003,cho2021,skinner2022}, or could be caused by instrumental systematic effects \citep{murphy2023}.
The planet-to-star radii ratios obtained in the Spitzer/IRAC passbands are consistent with the JWST/NIRSpec measurements within 1$\sigma$. The redder passbands have larger radii ratios at the 12$\sigma$ level for JWST/NIRSpec. The Spitzer/IRAC measurements present a similar, but smaller, trend, that is not significant due to much larger error bars.

\subsubsection{Spitzer/IRAC}
Two other eclipses of WASP-121 b were observed with each of the Spitzer/IRAC channels in 2017, as part of the program ID 13044 (PI: Drake Deming). \cite{garhart2020} reported lower dayside temperatures of 2490$\pm$77\,K (3.6\,$\mu$m) and 2562$\pm$66\,K (4.5\,$\mu$m) for WASP-121 b, based on those eclipse observations. The corresponding planet-to-star flux ratios reported by \cite{garhart2020} are (3.685$\pm$0.114)$\times$10$^{-5}$ and (4.684$\pm$0.121)$\times$10$^{-5}$, which are smaller than our measurements by 545 and 406 ppm, respectively. These differences could be caused by the phase-blend effect \citep{martin-lagarde2020}, that was likely neglected by \cite{garhart2020}. We estimated this effect for the 4.5\,$\mu$m eclipse using \texttt{ExoTETHyS.BOATS}, assuming the dayside and nightside temperatures reported in Table \ref{tab:posteriors} and the 8.5-hour duration of the Spitzer AORs. Indeed, the resulting phase-blend bias was -391 ppm, which is very similar to the discrepancy between the flux ratio reported by \cite{garhart2020} and our value (-406 ppm).

\subsubsection{HST/WFC3}

\cite{mikal-evans2022} analysed two phase curves of WASP-121 b observed with HST/WFC3 using G141 grism, which are spectrally resolved over 1.1-1.7\,$\mu$m. They reported dayside and nightside spectra with significant deviations from blackbody spectra, which they attributed to emission and absorption of H$^-$ and H$_2$O. Nonetheless, they adopted the brightness temperatures derived from the blackbody fits for the dayside and nightside hemispheres to estimate the Bond albedo and circulation efficiency of WASP-121 b atmosphere, finding $A_{\mathrm{b}} = 0.14 \pm 0.08$ and  $\varepsilon = 0.29 \pm 0.02$. We calculated the corresponding brightness temperatures to be $T_{\mathrm{day}}=2760 \pm 100 \,\mathrm{K}$ and $T_{\mathrm{night}}=1665 \pm 65 \,\mathrm{K}$ (not reported by \citealp{mikal-evans2022}). These HST/WFC3 observations suggest significantly lower Bond albedo, higher circulation efficiency and nightside temperatures than those that we obtained from Spitzer/IRAC observations. However, given the different wavelength ranges probed by HST and Spitzer observations, these apparent discrepancies do not necessarily reveal physical odds, but rather the limits of an oversimplified model behind these calculations. 

\cite{keating2017} pointed out that HST/WFC3 brightness temperatures can be overestimated due to neglecting the reflected star light component, namely, interpreting the observed flux from the planet dayside as pure emission.
We calculated the reflected light component integrated over the HST/WFC3 G141 passband to be $\sim$92 ppm in eclipse, assuming a geometric albedo of 0.32 (equal to the Bond albedo from Spitzer/IRAC channel 2). Neglecting this component could bias the inferred dayside temperatures by about +50\,K, but it should not affect the nightside temperature estimates.

\cite{changeat2022} performed joint retrievals on a suite of emission and transmission spectra of WASP-121 b, based on Spitzer/IRAC and HST/WFC3 observations. They retrieved atmospheric temperatures, weighted by the contribution function, of 2602$\pm$53\,K for the dayside and 1386$_{-366}^{+340}$\,K for the terminator. We note that the terminator temperature is not informed by the nightside spectrum and should be intermediate between the dayside and nightside temperatures.

Concerning the phase-curve maxima, \cite{mikal-evans2022} found modest offsets ahead of mid-eclipse for HST/WFC3 broadband and spectroscopic light curves. Their posterior median for the broadband light-curve fit is $\Delta \phi_{\mathrm{day}}^{\mathrm{MAX}} \sim -6^{\circ}$, in the opposite direction of our Spitzer/IRAC measurements.

\subsubsection{TESS}
The optical phase curve of WASP-121 b, as obtained from TESS data, has an amplitude of $\sim$400-500 ppm \citep{bourrier2020,daylan2021}. The former study found two solutions consistent with purely reflected starlight, leading to geometric albedo of $\sim$0.37, or pure thermal emission with $T_{\mathrm{day}}=2870 \pm 50 \, \mathrm{K}$ and $T_{\mathrm{night}} < 2200 \, \mathrm{K}$ (3$\sigma$). The latter assumed pure thermal emission with a  different stellar template from \cite{stassun2019}, leading to $T_{\mathrm{day}}=3012_{-42}^{+40} \, \mathrm{K}$ and $T_{\mathrm{night}}=2022_{-602}^{+44} \, \mathrm{K}$. From the second set of results \citep{daylan2021}, we calculated $A_{\mathrm{b}}=0.05_{-0.22}^{+0.18}$ and $\varepsilon=0.40_{-0.28}^{+0.17}$. We note that infrared observations provide tighter constraints on the atmospheric thermal properties, thanks to their larger phase-curve amplitudes and less reflection. Even neglecting reflection, our Spitzer/IRAC error bars on $A_{\mathrm{b}}$ and $\varepsilon$ are 2-8 times smaller than TESS ones.

\subsection{Comparison with other planets}

\begin{table*}[]
\renewcommand*{\arraystretch}{1.4}
\caption{Thermal phase-curve parameters of UHJs observed with Spitzer/IRAC.}
\centering
\begin{tabular}{ccccccc}
\hline
Planet & Wavelength [$\mu$m] & $T_{\mathrm{day}}$ [K] & $T_{\mathrm{night}}$ [K] & $\Delta \phi_{\mathrm{day}}^{\mathrm{MAX}}$ [deg] & $A_{\mathrm{b}}$ [] & $\varepsilon$ [] \\
\hline
\multirow{2}{*}{WASP-19 b\tablefootmark{i}} & 3.6 & 2384$_{-57}^{+41}$ & 890$_{-890}^{+280}$ & -10.5$\pm$4.0 & 0.35$_{-0.09}^{+0.08}$ & 0.05$_{-0.05}^{+0.09}$ \\
 & 4.5 & 2357$\pm$64 & 1130$_{-130}^{+240}$ & -12.9$\pm$3.6 & 0.33$_{-0.11}^{+0.09}$ & 0.13$_{-0.05}^{+0.12}$ \\
\hline
\multirow{2}{*}{HAT-P-7 b\tablefootmark{i}} & 3.6 & 2632$\pm$77 & $<$1360 (2$\sigma$) & 6.8$\pm$7.5 & 0.18$_{-0.21}^{+0.16}$ & $<$0.02 (2$\sigma$) \\
 & 4.5 & 2682$\pm$49 & 1710$\pm$180 & 4.1$\pm$7.5 & -0.13$_{-0.29}^{+0.22}$ & 0.35$_{-0.11}^{+0.12}$ \\
 \hline
\multirow{2}{*}{WASP-76 b\tablefootmark{ii}} & 3.6 & 2471$\pm$27 & 1518$\pm$61 & -0.68$\pm$0.48 & 0.24$_{-0.08}^{+0.07}$ & 0.31$\pm$0.04 \\
 & 4.5 & 2699$\pm$32 & 1259$\pm$44 & -0.67$\pm$0.20 & 0.05$\pm$0.09 & 0.117$_{-0.015}^{+0.017}$ \\
\hline
\multirow{2}{*}{WASP-33 b\tablefootmark{iii}} & 3.6 & 3082$\pm$92 & 1952$_{-134}^{+125}$ & -12.8$\pm$5.8 & 0.25$_{-0.10}^{+0.09}$ & 0.34$\pm$0.06 \\
 & 4.5 & 3209$_{-87}^{+89}$ & 1498$_{-118}^{+114}$ & -19.8$\pm$3.0 & 0.25$_{-0.09}^{+0.08}$ & 0.12$\pm$0.03 \\
\hline
\multirow{2}{*}{\textbf{WASP-121 b}\tablefootmark{iv}} & 3.6 & 2670$_{-40}^{+55}$ & 710$_{-710}^{+270}$ & 5.9$\pm$1.6 & 0.37$_{-0.09}^{+0.07}$ & 0.013$_{-0.013}^{+0.034}$ \\
 & 4.5 & 2700$_{-50}^{+70}$ & 1130$_{-160}^{+130}$ & 5.0$_{-3.4}^{+3.1}$ & 0.32$_{-0.10}^{+0.08}$ & 0.077$_{-0.034}^{+0.040}$ \\
 \hline
\multirow{2}{*}{KELT-1 b\tablefootmark{v}} & 3.6 & 2988$\pm$60 & 1173$_{-130}^{+175}$ & -28.4$\pm$3.5 & 0.09$_{-0.10}^{+0.09}$ & 0.06$_{-0.02}^{+0.04}$ \\
 & 4.5 & 2902$\pm$74 & 1053$_{-161}^{+230}$ & -18.6$\pm$5.2 & 0.19$_{-0.10}^{+0.09}$ & 0.05$_{-0.02}^{+0.05}$ \\
 \hline
\multirow{2}{*}{WASP-103 b\tablefootmark{vi}} & 3.6 & 2995$\pm$159 & 1523$\pm$153 & -2.0$\pm$0.7 & 0.15$_{-0.21}^{+0.17}$ & 0.16$_{-0.06}^{+0.08}$ \\
 & 4.5 & 3154$\pm$99 & 1288$\pm$118 & -1.0$\pm$0.4 & 0.02$_{-0.18}^{+0.15}$ & 0.07$_{-0.02}^{+0.03}$ \\
 \hline
\multirow{4}{*}{WASP-12 b\tablefootmark{vii}} & 3.6 & 2744$\pm$48 & 1510$\pm$210 & N.A. & 0.38$_{-0.12}^{+0.10}$ & 0.21$_{-0.09}^{+0.12}$ \\
 & 3.6 & 2813$\pm$48 & 1760$\pm$97 & N.A. & 0.26$_{-0.13}^{+0.11}$ & 0.33$\pm$0.06 \\
 & 4.5 & 2989$\pm$66 & 790$\pm$150 & N.A. & 0.24$_{-0.14}^{+0.12}$ & 0.013$_{-0.007}^{+0.013}$ \\
  & 4.5 & 2854$\pm$74 & 1340$\pm$180 & N.A. & 0.32$_{-0.13}^{+0.11}$ & 0.12$_{-0.05}^{+0.07}$ \\
 \hline
KELT-9 b\tablefootmark{viii} & 4.5 & 4566$_{-136}^{+140}$ & 2556$_{-97}^{+101}$ & -18.7$_{-2.3}^{+2.1}$ & 0.29$_{-0.18}^{+0.14}$ & 0.23$\pm$0.04 \\
\hline
\end{tabular}
\tablefoot{The brightness temperatures and hot-spot offsets were extracted from the literature, and used to calculate the Bond albedo and circulation efficiency.
References:
\tablefoottext{i}{\cite{wong2016}};
\tablefoottext{ii}{\cite{may2021}};
\tablefoottext{iii}{\cite{zhang2018}};
\tablefoottext{iv}{This work};
\tablefoottext{v}{\cite{beatty2019}};
\tablefoottext{vi}{\cite{kreidberg2018}};
\tablefoottext{vii}{\cite{bell2019}};
\tablefoottext{viii}{\cite{mansfield2020}}.
}
\label{tab:Ab_eps_comparison}
\end{table*}

WASP-121 b belongs the class of UHJs, i.e., gas giants with dayside temperature $\gtrsim$2200\,K \citep{bell2018}. This temperature is above the condensation threshold of most species, except highly refractory ones such as Al and Ti \citep{wakeford2017}. For this reason, we may expect cloud-free dayside in UHJs \citep{helling2021}. The lack of a reflecting cloud layer should also imply low geometric and Bond albedo ($\lesssim$0.2), as confirmed by optical to near-infrared eclipse measurements \citep{mallonn2019}. We estimated a higher Bond albedo of $\gtrsim$0.3 from the Spitzer/IRAC phase curves of WASP-121 b at 3.6 and 4.5\,$\mu$m, although their posteriors are consistent with $A_{\mathrm{b}}=0.2$ within 2$\sigma$. \cite{schwartz2015} highlighted a common trend of measuring systematically higher Bond albedos from thermal phase curves of gas giants ($A_{\mathrm{b}} \sim 0.35$) compared to the geometric albedos inferred from visible eclipses ($A_{\mathrm{g}} \sim 0.1$). This trend holds, with a few exceptions, for more recent observations of UHJs (see Table \ref{tab:Ab_eps_comparison}).

The thermal phase curves of UHJs typically have large amplitudes, corresponding to strong day-night contrasts, and small hot-spot offsets. These properties indicate inefficient heat redistribution. We derived $\varepsilon \lesssim 0.1$ for most UHJs, the lowest values were obtained for WASP-121 b (see Table \ref{tab:Ab_eps_comparison}). There are no evident trends between the irradiation temperatures of UHJs and the observed thermal phase-curve parameters.

\section{Conclusions}
\label{sec:conclusions}

We analysed, for the first time, two thermal phase curves of WASP-121 b taken with Spitzer. Despite these datasets being affected by stronger than usual instrumental systematic effects, we obtained meaningful information on the exoplanet atmosphere. The measured brightness temperatures and transit depths are consistent within 1$\sigma$ with those obtained from much more precise JWST observations in similar passbands. We estimated the Bond albedo and circulation efficiency of the WASP-121 b atmosphere, which are similar to those of other UHJs. However, we measured unusual westward hot-spot offsets, which are significantly different from the JWST measurements. These discrepancies may hint at atmospheric variability or instrumental systematic effects. We further explored the possible thermal profiles using phase-curve retrievals, which are coupled with chemistry. Our analysis confirms the validity of Spitzer phase-curves to infer exoplanet atmospheric properties. More precise, spectrally resolved observations, such as those obtained with JWST, will enable us to better understand their complex behaviour.

\begin{acknowledgements}
This work is based on archival data obtained with the Spitzer Space Telescope, which was operated by the Jet Propulsion Laboratory, California Institute of Technology under a contract with NASA.
G. M. has received funding from the European Union's Horizon 2020 research and innovation programme under the Marie Sk\l{}odowska-Curie grant agreement No. 895525, and from the Ariel Postdoctoral Fellowship program of the Swedish National Space Agency (SNSA). 
Q. C. is funded by the European Space Agency under the 2022 ESA Research Fellowship Program.
We acknowledge the availability and support from the High Performance Computing platforms (HPC) from the Simons Foundation (Flatiron), DIRAC and OzSTAR, which provided the computing resources necessary to perform this work. This work was performed using the Cambridge Service for Data Driven Discovery (CSD3), part of which is operated by the University of Cambridge Research Computing on behalf of the STFC DiRAC HPC Facility (www.dirac.ac.uk). The DiRAC component of CSD3 was funded by BEIS capital funding via STFC capital grants ST/P002307/1 and ST/R002452/1 and STFC operations grant ST/R00689X/1. DiRAC is part of the National e-Infrastructure. Additionally, this work utilised the OzSTAR national facility at Swinburne University of Technology. The OzSTAR program receives funding in part from the Astronomy National Collaborative Research Infrastructure Strategy (NCRIS) allocation provided by the Australian Government.
\end{acknowledgements}

%
%

\bibliographystyle{aa_url} 
\bibliography{biblio} 

\onecolumn
\begin{appendix}

\section{Corner plots}
\label{app:cornerplots}
\begin{figure}[!h]
\includegraphics[width=0.95\hsize]{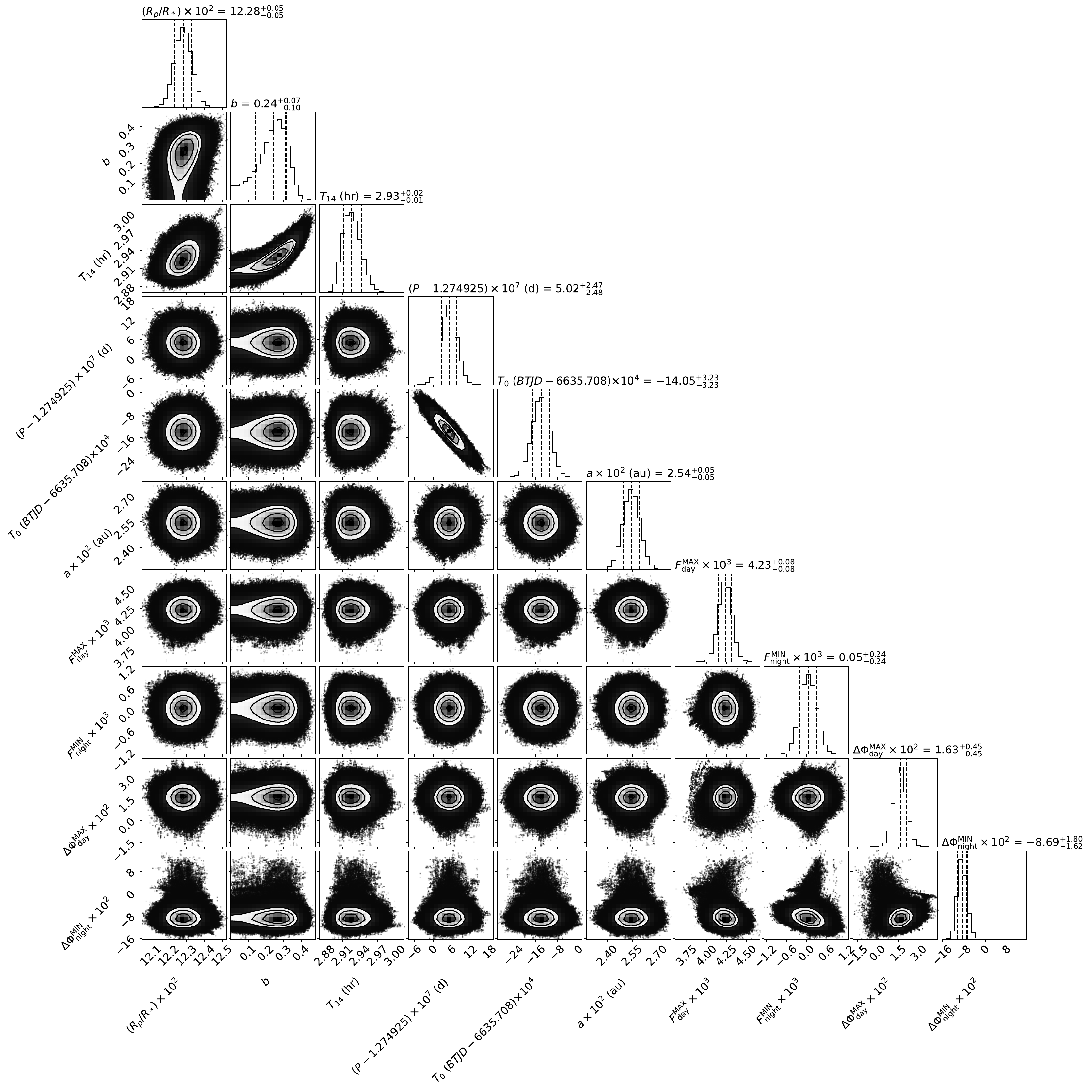}
\caption{Posterior distributions of the light-curve fitting performed on the Spitzer/IRAC channel 1 data.}
\label{fig:corner1}
\end{figure}

\begin{figure}
\includegraphics[width=0.95\hsize]{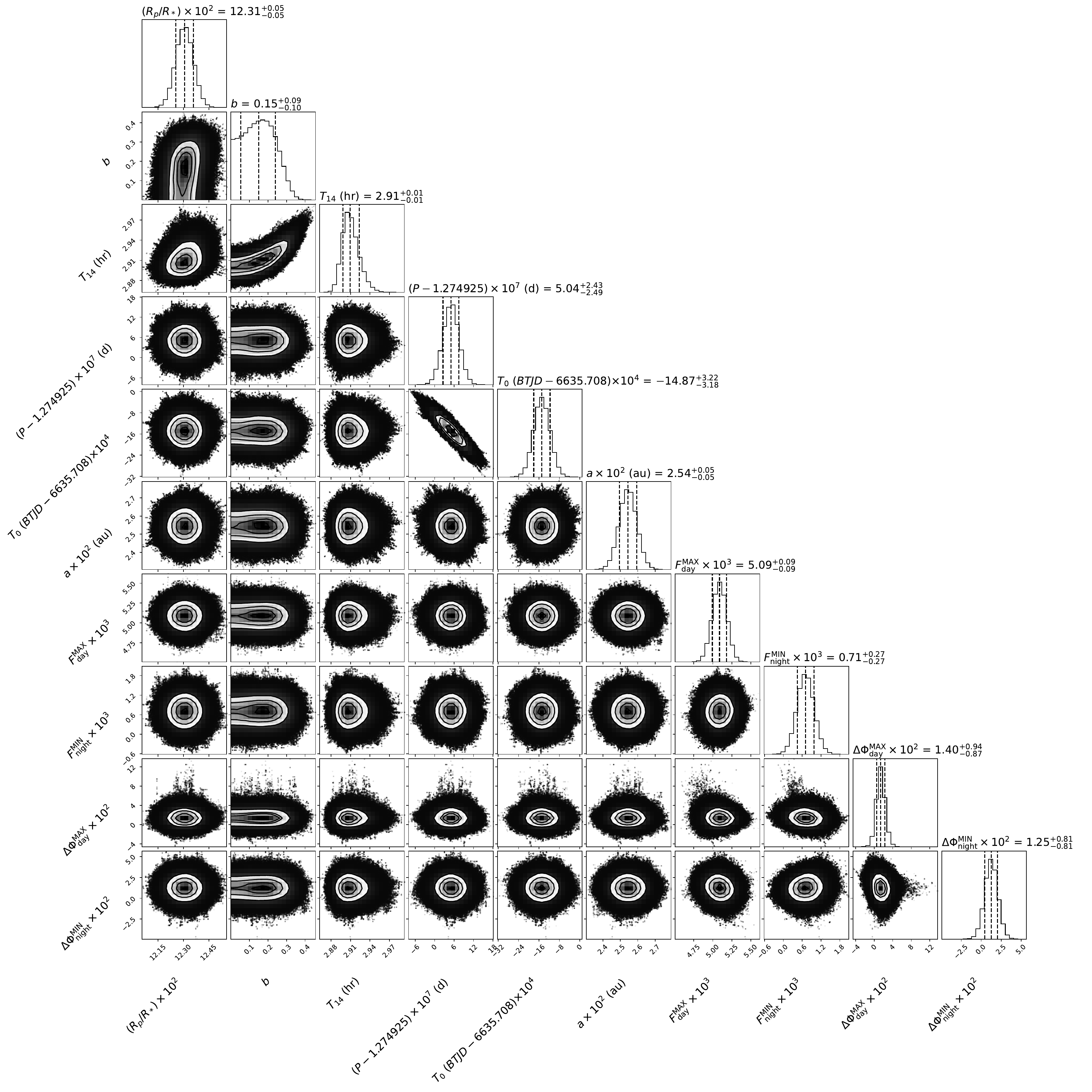}
\caption{Posterior distributions of the light-curve fitting performed on the Spitzer/IRAC channel 2 data.}
\label{fig:corner2}
\end{figure}

\begin{figure}
\includegraphics[width=0.95\hsize]{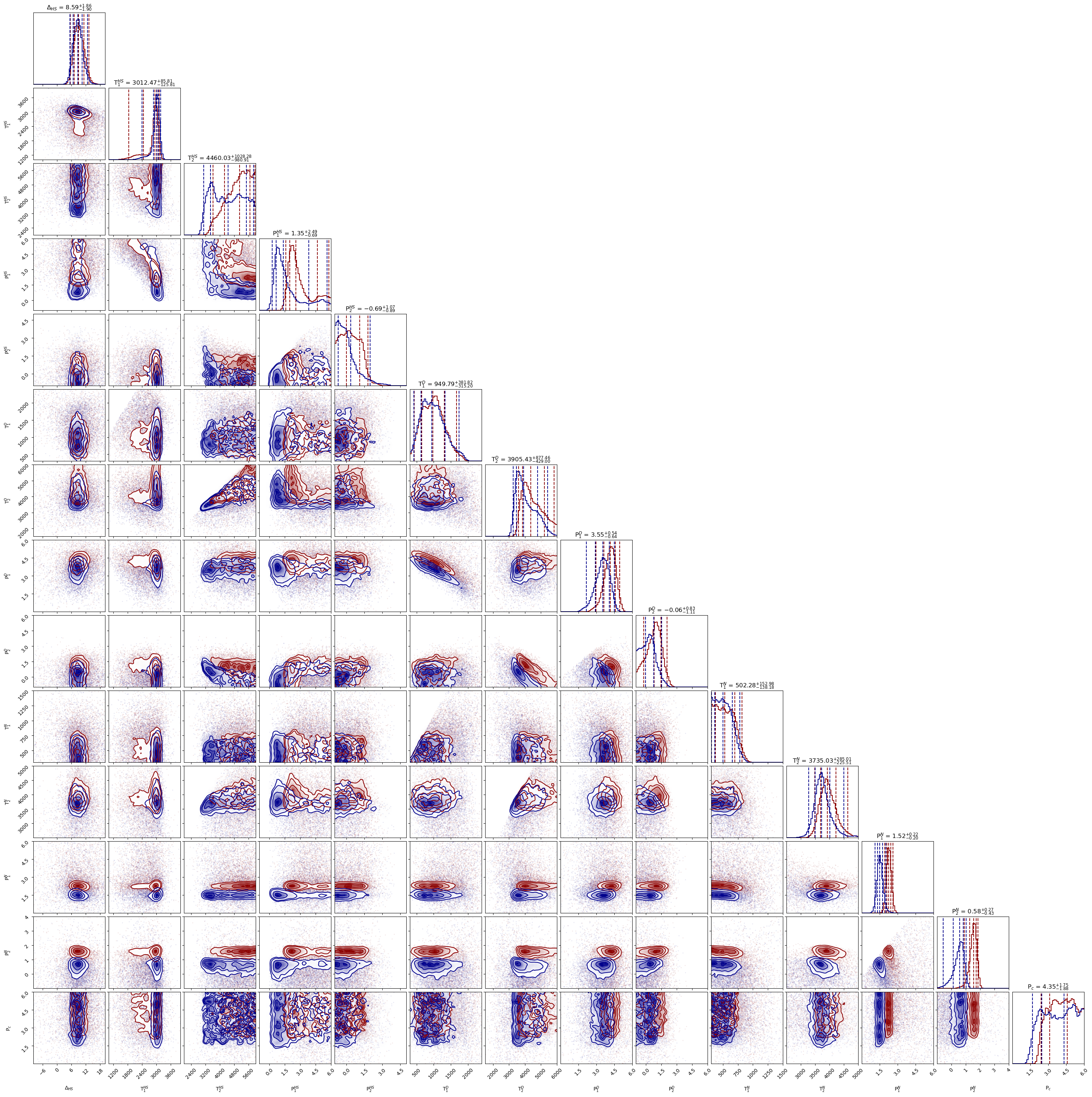}
\caption{Posterior distributions of the atmospheric retrievals performed on the Spitzer data. Blue: retrieval with $Z_{\mathrm{p}} = 1 \, Z_\odot$; Red: Retrieval with $Z_{\mathrm{p}} = 10 \, Z_\odot$. $\Delta_{HS}$ is the hot-spot offset of the mode. HS: Hot-spot. D: Day. N: Night. P$_c$ is the top pressure of the cloud deck. }
\label{fig:corner_atmosphere}
\end{figure}

Figures \ref{fig:corner1} and \ref{fig:corner2} show the corner plots with the posterior distributions of astrophysical parameters obtained from the 3.6\,$\mu$m and 4.5\,$\mu$m observations, respectively.
Figure \ref{fig:corner_atmosphere} shows the corner plots for the atmospheric parameters retrieved from both observations assuming $Z_{\mathrm{p}} = 1 \, Z_\odot$ and $Z_{\mathrm{p}} = 10 \, Z_\odot$.

\end{appendix}

\end{document}